\documentstyle[multicol,aps,epsf,preprint]{revtex}
\begin{document}
\draft \title{Simulations of liquid crystals in Poiseuille flow} 
\author{Colin Denniston$^{1}$, Enzo Orlandini$^{2}$, and J.M. Yeomans$^{1}$}
\address{$^{1}$ Dept. of Physics, Theoretical Physics,
University of Oxford, 1 Keble Road, Oxford OX1 3NP}
\address{$^{2}$ INFM-Dipartimento di Fisica, Universit\`a di
  Padova, 1-35131 Padova, Italy}
\date{\today} \maketitle

\begin{abstract}
Lattice Boltzmann simulations are used to explore the behavior of
liquid crystals subject to Poiseuille flow. In the nematic regime  
at low shear rates we find two possible steady state configurations of
the director field.  The selected state depends on both
the shear rate and the history of the sample.  For both director
configurations there is clear evidence
of shear-thinning, a decrease in the viscosity with increasing shear rate.
Moreover, at very high shear rates or when the order parameter is
large, the system transforms to a log-rolling state with boundary
layers that may exhibit oscillatory behavior.
\end{abstract}

\renewcommand{\theequation}{\thesection.\arabic{equation}}
\renewcommand{\baselinestretch}{1.2} \def\xbar{1/x} \def\ra{\rangle}
\def\la{\langle} \def\Pr{{\it Proof: }} \def\qed{$\Box$} \def\cF{{\cal
F}} \def\cFo{{\cal F}^o} \def\pan{\par\noindent}
\def\pa{\partial_{\alpha}} \def\pb{\partial_{\beta}}
\def\pe{\partial_{\eta}} \def\pt{\partial_t} \def\eia{e_{i\alpha}}
\def\eib{e_{i\beta}} \def\eie{e_{i\eta}} \def\geq{g_i^{eq}}
\def\feq{f_i^{eq}} \def\dt{\Delta t} \def\ua{u_{\alpha}}
\def\ue{u_{\eta}} \def\Tea{T_{\eta\alpha}} \def\Pea{P_{\eta\alpha}}
%


\section{Introduction}
As suggested by their name, liquid crystals are fundamentally a liquid
and hence hydrodynamics can be very important to their behavior.
Moreover, because they comprise rod-like molecules the properties of 
their flow can be much richer than that seen
in simple fluids.  This is because the translational motion of the
fluid is coupled to the inner orientational motion of the molecules.
As a result the flow disturbs the
alignment.  Conversely changing the alignment will almost invariably
induce a flow.

This coupling can have important practical consequences in
applications of liquid crystals.  Back-flow can be an important
limiting factor in determining the rate at which a
liquid crystal display device can be switched \cite{Sharp}.  The
ordering induced by a shear flow could potentially be beneficial if
understood and controlled or can be the origin of mechanical
instabilities under processing conditions \cite{Process,Process2}.

In addition to technologically relevant issues, there are also a
number of fundamental scientific questions which can be addressed in
the context of liquid crystals.  It is known that shear flow can
induce a phase transition from the isotropic (disordered) to nematic
(liquid crystalline ordered) phase\cite{O92,M97,O99}.  
The resulting non-equilibrium phase
diagram includes a line of first-order transitions ending in a
critical point.  Techniques for examining and classifying such
non-equilibrium phase transitions are still in need of further
testing and development.  In addition, a large number of structured
fluids, like a liquid crystal, have order parameters that have tensor
characteristics.  Methods for studying such systems are still somewhat
lacking and need to be explored.

Although the effects of hydrodynamics are well known, it is usually
very difficult to fully incorporate these effects into a calculation.  
This has been due to the
complexity of the nine coupled non-linear partial differential
equations which describe the system (the order parameter
is a symmetric traceless tensor with five components, the
Navier-Stokes equation has three components, and the continuity
equation is scalar).  While the final results may appear simple it can
often be an imposing task to derive them analytically from the
underlying equations.  From a numerical viewpoint accurate simulations
can be difficult as the solutions can contain topological defects
requiring resolution of length scales much smaller than those
associated with hydrodynamic flows.

In an attempt to overcome these problems we have developed a modified
lattice Boltzmann algorithm to simulate liquid crystals.  
The lattice Boltzmann algorithm, which includes hydrodynamics, makes use of a
physical analogue (from statistical mechanics) to map the partial
differential equations onto equations for the evolution of two
probability distributions. Moments of these distributions are then
related to the physical variables of interest.  Among the advantages
of this reformulation is that the resulting equations are much simpler
to model and possess greater numerical stability than, say, a
traditional finite-difference scheme applied to the original equations.   

In this paper we apply the lattice Boltzmann method to the study of
Poiseuille flow in liquid crystals.  The paper is organized as
follows.  First, we describe the model of liquid crystal hydrodynamics
proposed by Beris and Edwards\cite{BE}. This contains both the Erickson-Leslie
and the Doi models as limiting cases.  Second, we outline the main
features of  the lattice Boltzmann algorithm used to simulate the
model.  Results are presented to show the effect of the flow on the
director orientation and how this in turn leads to both shear thinning
and hysteresis. We also examine the transition to the log-rolling state.

\section{The equations of motion for liquid crystal hydrodynamics}

We follow the formulation of liquid crystal hydrodynamics described by
Beris and Edwards\cite{BE}. The continuum equations of motion are written in
terms of a tensor order parameter ${\bf Q}$. The advantage of this
approach is that it includes both the isotropic (${\bf Q}=0$) and the
nematic (${\bf Q} \neq 0$) phases and allows an order parameter of
variable magnitude within the latter. Hence it is possible to explore
the effect of flow on the phase transition between the two states.
Moreover the hydrodynamics of topological defects (point defects in
two dimensions) are naturally included in the equations. The
Beris--Edwards equations reduce to the Ericksen--Leslie formulation
\cite{deGennes93} of nematodynamics for a uniaxial nematic in the
absence of defects. 

We describe the equilibrium properties of the liquid crystal by the
Landau--de Gennes free energy\cite{deGennes93}
\begin{equation}
{\cal F}=\int d^3 r \left\{ \frac{a}{2} 
  Q_{\alpha \beta}^2 - \frac{b}{3} Q_{\alpha \beta}Q_{\beta \gamma}
Q_{\gamma \alpha}+ \frac{c}{4}
  (Q_{\alpha \beta}^2)^2 + \frac{\kappa}{2} (\partial_\alpha Q_{\beta
  \lambda})^2
  \right\},
\label{free}
\end{equation}
where Greek indices represent Cartesian directions and a sum over
repeated indices is assumed.
This free energy describes a first-order transition from the isotropic
to the nematic phase.
Note that, for simplicity, we are working within the one elastic constant
approximation. Three elastic constants are needed to fully
characterize the nematic phase\cite{deGennes93}.

The equation of motion for the nematic order parameter is
\begin{equation}
(\partial_t+{\vec u}\cdot{\bf \nabla}){\bf Q}-{\bf S}({\bf W},{\bf
  Q})= \Gamma {\bf H}
\label{Qevolution}
\end{equation}
where $\Gamma$ is a collective rotational diffusion constant.
The first term on the left-hand side of equation (\ref{Qevolution})
is the material derivative describing the usual time dependence of a
quantity advected by a fluid with velocity ${\vec u}$. This is
generalized by a second term 
\begin{eqnarray}
{\bf S}({\bf W},{\bf Q})
&=&(\xi{\bf D}+{\bf \Omega})({\bf Q}+{\bf I}/3)+({\bf Q}+
{\bf I}/3)(\xi{\bf D}-{\bf \Omega})\nonumber\\
& & -2\xi({\bf Q}+{\bf I}/3){\mbox{Tr}}({\bf Q}{\bf W})
\end{eqnarray}
where ${\bf D}=({\bf W}+{\bf W}^T)/2$ and
${\bf \Omega}=({\bf W}-{\bf W}^T)/2$
are the symmetric part and the anti-symmetric part respectively of the
velocity gradient tensor $W_{\alpha\beta}=\partial_\beta u_\alpha$.
${\bf S}({\bf W},{\bf Q})$  appears in the equation of motion because
the rod-like shape of the molecules means that
the order parameter distribution can be both rotated and stretched by
flow gradients. $\xi$ is a constant which depends on the molecular
details of a given liquid crystal.

The term on the right-hand side of equation \ref{Qevolution}
describes the relaxation of the order parameter towards the minimum of
the free energy. The molecular field ${\bf H}$ which provides the
driving
force is related to the derivative of the free energy by
\begin{eqnarray}
{\bf H}&=& -\left\{{\delta {\cal F} \over \delta Q}-({\bf
    I}/3) Tr{\delta {\cal F} \over \delta Q}\right\} 
\label{H(Q)}\\ 
&=&    -a {\bf Q}+ b \left({\bf Q^2}-({\bf
    I}/3)Tr{\bf Q^2}\right)- c  {\bf Q}Tr{\bf Q^2}+\kappa \nabla^2
    {\bf Q}.
\nonumber
\end{eqnarray}  

The fluid, of density $\rho$ obeys the continuity
\begin{equation}
\partial_t \rho + \partial_{\alpha} \rho u_{\alpha} =0
\label{continuity}
\end{equation}
and the Navier-Stokes equations 
\begin{eqnarray}
& & \rho\partial_t u_\alpha+\rho u_\beta \partial_\beta
u_\alpha=\partial_\beta \tau_{\alpha\beta}+\partial_\beta
\sigma_{\alpha\beta} \label{NS}\\
& & \qquad+{\rho \tau_f \over
3}(\partial_\beta((\delta_{\alpha\beta}-3\partial_\rho
P_{0})\partial_\gamma u_\gamma+\partial_\alpha
u_\beta+\partial_\beta u_\alpha).
\nonumber
\end{eqnarray}
where $\tau_f$ is related to the viscosity and $P_0$ is the pressure,
\begin{equation}
P_0=\rho T-\frac{\kappa}{2} (\nabla {\bf Q})^2.
\end{equation}
The form of this equation is not dissimilar to that for a simple
fluid. However the details of the stress tensor reflect the additional
complications of liquid crystal hydrodynamics. 
There is a symmetric contribution
\begin{eqnarray}
\sigma_{\alpha\beta} &=&-P_0 \delta_{\alpha \beta}
-\xi H_{\alpha\gamma}(Q_{\gamma\beta}+{1\over
  3}\delta_{\gamma\beta}) \nonumber\\
& & \, -\xi (Q_{\alpha\gamma}+{1\over
  3}\delta_{\alpha\gamma})H_{\gamma\beta}
 +2\xi
(Q_{\alpha\beta}+{1\over 3}\delta_{\alpha\beta})Q_{\gamma\epsilon}
H_{\gamma\epsilon}-\partial_\beta Q_{\gamma\nu} {\delta
{\cal F}\over \delta\partial_\alpha Q_{\gamma\nu}}
\label{BEstress}
\end{eqnarray}
and an antisymmetric contribution
\begin{equation}
 \tau_{\alpha \beta} = Q_{\alpha \gamma} H_{\gamma \beta}-
 H_{\alpha \gamma} Q_{\gamma \beta}.
\label{as}
\end{equation}
A detailed account of the theories of liquid crystal hydrodynamics
can be found in the book by Beris and Edwards \cite{BE}.

\section{A lattice Boltzmann algorithm for liquid crystal
hydrodynamics}

We now define a lattice Boltzmann algorithm which solves the
hydrodynamic equations of motion of a liquid crystal 
(\ref{Qevolution}), (\ref{continuity}), and (\ref{NS}). 
Lattice Boltzmann algorithms are defined in
terms of a set of continuous variables, usefully termed partial
distribution functions, which move on a lattice in discrete space and
time. For a simple fluid a
single set of partial distribution functions which sum on each site to
give the density is needed. For liquid crystal hydrodynamics this must be
supplemented by a second set, which are tensor variables, and which
are related to the tensor order parameter ${\bf Q}$.

We define two distribution functions, the scalars $f_i (\vec{x})$ and
the symmetric traceless tensors ${\bf G}_i (\vec{x})$ on each lattice
site $\vec{x}$. Each $f_i$, ${\bf G}_i$ is associated with a lattice
vector ${\vec e}_i$. We choose a nine-velocity model on a square
lattice with velocity vectors ${\vec e}_i=(\pm 1,0),(0,\pm 1), (\pm
\sqrt{2}, \pm
\sqrt{2}), (0,0)$. Physical variables are defined as moments of the
distribution function
\begin{equation}
\rho=\sum_i f_i, \qquad \rho u_\alpha = \sum_i f_i  e_{i\alpha},
\qquad {\bf Q} = \sum_i {\bf G}_i.
\label{eq1}
\end{equation} 

The distribution functions evolve in a time step $\Delta t$ according
to
\begin{eqnarray}
& &f_i({\vec x}+{\vec e}_i \Delta t,t+\Delta t)-f_i({\vec x},t)=\label{eq2}\\
& & \qquad \frac{\Delta t}{2} \left[{\cal C}_{fi}({\vec x},t,\left\{f_i
\right\})+ {\cal C}_{fi}({\vec x}+{\vec e}_i \Delta
t,t+\Delta
t,\left\{f_i^*\right\})\right],
\nonumber
\end{eqnarray}
\begin{eqnarray}
& &{\bf G}_i({\vec x}+{\vec e}_i \Delta t,t+\Delta t)-{\bf G}_i({\vec
x},t)=\label{eq3}\\
& & \qquad \frac{\Delta t}{2}\left[ {\cal C}_{{\bf G}i}({\vec
x},t,\left\{{\bf G}_i \right\})+
                {\cal C}_{{\bf G}i}({\vec x}+{\vec e}_i \Delta
                t,t+\Delta t,\left\{{\bf G}_i^*\right\})\right]
\nonumber
\end{eqnarray}
The left-hand side of these equations represents free streaming with
velocity ${\vec e}_i$, while the right-hand side is a
collision step which allows the distribution to relax towards
equilibrium. $f_i^*$ and ${\bf G}_i^*$ are first order approximations
to 
$f_i({\vec x}+{\vec e}_i \Delta t,t+\Delta t)$ and ${\bf G}_i({\vec x}+{\vec
e}_i \Delta t,t+\Delta t)$
respectively. They are obtained from equations
(\ref{eq2}) and (\ref{eq3}) but with $f_i^*$ and ${\bf G}_i^*$ set to
$f_i$ and ${\bf G}_i$.
Discretizing in this way, which is similar to a predictor-corrector 
scheme, has the advantages that lattice viscosity terms are eliminated
to second order and that the stability of the scheme is improved.

The collision operators are taken to have the form of a single
relaxation time Boltzmann equation, together with a forcing term,
\begin{eqnarray}
& & {\cal C}_{fi}({\vec x},t,\left\{f_i \right\})=\label{eq4}\\
& & \qquad -\frac{1}{\tau_f}(f_i({\vec x},t)-f_i^{eq}({\vec x},t,\left\{f_i
\right\}))
+p_i({\vec x},t,\left\{f_i \right\}),
\nonumber\\
& &{\cal C}_{{\bf G}i}({\vec x},t,\left\{{\bf G}_i
\right\})=\label{eq5}\\
& & \qquad -\frac{1}{\tau_{\bf G}}({\bf G}_i({\vec x},t)-{\bf
G}_i^{eq}({\vec x},t,\left\{{\bf G}_i \right\}))
+{\bf M}_i({\vec x},t,\left\{{\bf G}_i \right\}).
\nonumber
\end{eqnarray}

The form of the equations of motion and of thermodynamic equilibrium
follow from the choice of the moments of the equilibrium distributions
$f^{eq}_i$ and ${\bf G}^{eq}_i$ and the driving terms $p_i$ and
${\bf M}_i$. $f_i^{eq}$ is constrained by
\begin{equation}
\sum_i f_i^{eq} = \rho,\quad \sum_i f_i^{eq} e_{i \alpha} = \rho
u_{\alpha}, \quad
\sum_i f_i^{eq} e_{i\alpha}e_{i\beta} = -\sigma_{\alpha\beta}+\rho
u_\alpha u_\beta
\label{eq6} 
\end{equation}
where the zeroth and first moments are chosen to impose conservation
of
mass and momentum. The second moment of $f^{eq}$ controls the
symmetric
part of the stress tensor, whereas the moments of $p_i$
\begin{equation}
\sum_i p_i = 0, \quad \sum_i p_i e_{i\alpha} = \partial_\beta
\tau_{\alpha\beta},\quad \sum_i p_i
e_{i\alpha}e_{i\beta} = 0
\label{eq7}
\end{equation}
impose the antisymmetric part of the stress tensor.
For the equilibrium of the order parameter distribution we choose
\begin{equation}
\sum_i {\bf G}_i^{eq} = {\bf Q},\quad \sum_i
{\bf G}_i^{eq} {e_{i\alpha}} = {\bf Q}{u_{\alpha}},
\quad
\sum_i {\bf G}_i^{eq}
e_{i\alpha}e_{i\beta} = {\bf Q} u_\alpha u_\beta .
\label{eq8}
\end{equation}
This ensures that the order parameter
is convected with the flow. Finally the evolution of the
order parameter towards the minimum of the free energy
is most conveniently modeled by choosing
\begin{equation}
\sum_i {\bf M}_i = \Gamma{\bf H}({\bf Q})
+{\bf S}({\bf W},{\bf Q}),
\quad \sum_i {\bf M}_i {e_{i\alpha}} = (\sum_i {\bf M}_i)
{u_{\alpha}}.
\label{eq9}
\end{equation}
Conditions (\ref{eq6})--(\ref{eq9})
can be satisfied as is usual in lattice Boltzmann
schemes by writing the equilibrium distribution functions and forcing
terms as polynomial expansions in the velocity \cite{C98}.

Taking the continuum limit of equations (\ref{eq2}) and (\ref{eq3}) and
performing a Chapman-Enskog expansion leads to the equations of motion
of liquid crystal hydrodynamics
(\ref{Qevolution}), (\ref{continuity}), and (\ref{NS}).

\section{Poiseuille flow}
We now consider Poiseuille flow in a liquid crystal.
Poiseuille flow is the flow between two non-slip boundaries at $y=0$ and
$y=h$,  driven by a pressure gradient or body force.
In a simple fluid this geometry leads to a parabolic profile of velocities
\cite{LandauLifshitz}
\begin{equation}
u_x=-\frac{1}{2\eta}(\rho b_f)y(y-h)
\label{upois}
\end{equation}
where $\eta$ is the viscosity, $(\rho b_f)$ is an applied body force, and
$u_x$ is the velocity along the channel.  A simple example is 
gravity-driven flow.

In an experiment the volumetric flow rate
\begin{equation}
Z=\int_0^h u_x dy= \frac{1}{12 \eta}(\rho b_f) h^3
\label{flowrate}
\end{equation}
can be measured as a 
function of the applied body force (or pressure gradient) to
obtain the viscosity $\eta$.  If the material being studied is
homogeneous, then
the viscosity should be a unique function of the effective shear rate,
$Z/h^2$.  This can be used as a working definition of a simple fluid.
If the fluid is not homogeneous, then the relation between $\eta$ and
the effective shear rate $Z/h^2$ will depend on geometrical factors
such as the channel width $h$.

In particular, a nematic liquid crystal has internal structure which
can lead to inhomogeneities.  The internal structure is strongly
affected by the presence of the flow field.  In a simple shear flow
the director in a nematic liquid crystal will try to align with the
flow direction at an angle $\theta$ given by
\begin{equation}
\xi \cos 2 \theta=\frac{3 q}{2+q},
\label{shearangle}
\end{equation}
where $q$ is the amplitude of the order parameter (equal to $3/2$
times the magnitude of the largest eigenvalue of ${\bf Q}$)\cite{BE}.
However
this condition is not compatible with fixed boundary conditions at
the walls of the channel and therefore the direction of the 
director field will be determined by a balance of
competing factors. In addition
note that because $\xi<1$ there will be no solution to equation
(\ref{shearangle}) as $q$ approaches one.  When no solution
is available the director tumbles in the flow.  An examination of the
different regimes of tumbling under shear
(but imposing a constant shear gradient across the sample rather 
than solving Navier-Stokes
equations with shear boundary conditions) is given in Ref.\cite{TsujiRey}.

In order to make it easier to compare our results to a wider array of
experimental situations, we will plot all quantities in dimensionless
numbers.  First, note that $1/\Gamma$ has units of viscosity and
$\kappa$ has units of force.  One can then form a velocity scale as
$\kappa \Gamma /h$.  Using this velocity scale one can construct the
Erickson number $Er= u_x  h/(\kappa \Gamma)$ (normally, in the
Erickson-Leslie theory, the Ericksen number is defined using a
viscosity.  However, it can be shown that $\Gamma$ sets a scale for
the contribution to the measured viscosity from the liquid crystal
order\cite{BE}, which is what we are interested in here).  The
Ericksen number for the flows we examine is in the range $10^2$ to
$10^4$.  For comparison, the Ericksen number for the flows in the
experiments described in Reference \cite{FF69} is in the range $10^2$
to $10^5$.  A useful time scale is $1/(p_0\Gamma)$ where $p_0$ is the
ambient pressure ($p_0=1\, atm$ in all our simulations).  All the
simulations are performed in two spatial dimensions, while the
order parameter is in three (i.e. the director can point out of the
plane). 

The remainder of the paper is organized as follows.  We first examine
the order parameter configurations which are possible in different
flow regimes.  Then the corresponding effective viscosity is measured
as a function of flow rate.  We find shear thinning, and that the
viscosity is not a unique function of the effective shear rate, in
good agreement with experiments.  Applying a scaling suggested by
Ericksen \cite{E69} we find that the data can be collapsed onto a single
curve with two branches.  The transition to the log-rolling state,
where the director points out of the shear plane, is then briefly
examined.  

\subsection{Order parameter profile}

\begin{figure}
\centerline{\epsfxsize=2.5in \epsffile{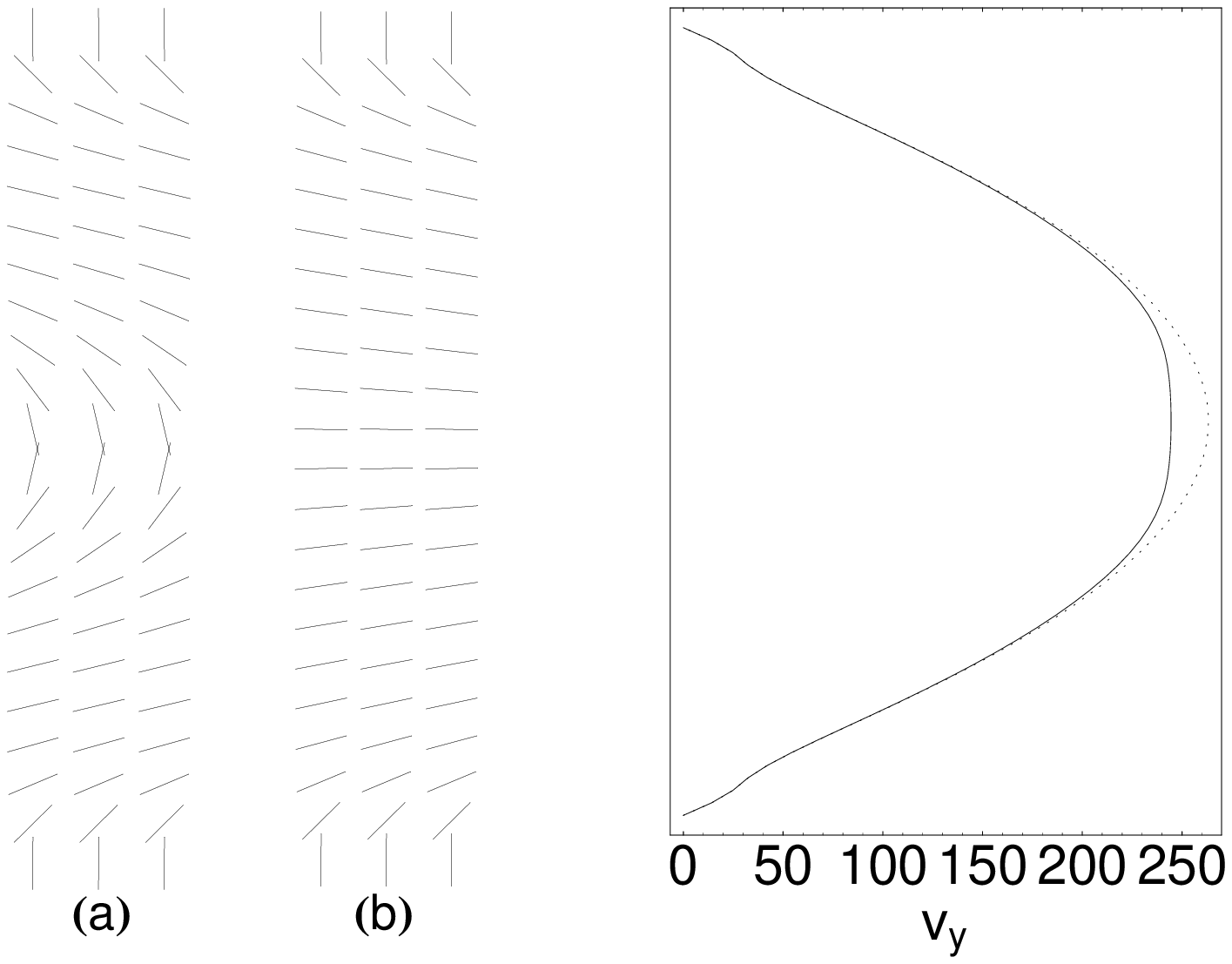}} \vskip
0.5true cm
\caption{Left: Two different director configurations for steady-state
  Poiseuille flow with the director in the plane.  The boundary conditions
  are such that the director is perpendicular ($\theta=90$) to the walls
  at the top and bottom of the figure and the flow is from left to
  right.  For clarity,
  only the director on every third lattice point is displayed.  Right:
  The corresponding fluid velocity $v_y$. The
  solid and dotted lines are for the director configurations (a) and
  (b) respectively. Velocities are scaled by $\kappa\Gamma/h$ to
  make them dimensionless.}
\label{SimplePois}
\end{figure}

Distortions in the nematic state are induced by
the laminar Poiseuille flow.  Two possible configurations of the director
field, for the case with strong normal anchoring at the boundaries, 
are shown in Figure~\ref{SimplePois}.
In  case (a) the system was initialized in the nematic phase with the
director uniform and perpendicular to the boundary before the flow was
turned on.  In case (b)  the system was kept in the isotropic
phase 
for the first 4000 time steps to establish
the flow and then quenched into the nematic phase.
Otherwise all parameters were identical.
It is clear that the two systems have chosen different
compatibility conditions at the center of the flow.  Configuration (a)
is described in the book by deGennes\cite{deGennes93}.  
Case (b) does not appear to have been previously considered in the context
of flow.  It is, however, very similar to the equilibrium director
configuration expected in a cylindrical tube with normal anchoring
boundary conditions.

To understand these configurations note that in both cases there are
 three distinct regions 
of the flow.  First, away from the walls and the center, the director is
oriented at the angle given by equation (~\ref{shearangle}).
Approaching the wall, the director must change its configuration to
match the boundary condition (normal anchoring).  The distance $e_1$ over
which this occurs is determined by a competition between the elastic
energy and the shear stress at the boundary and can be estimated as
\begin{equation}
e_1\sim (K/\eta s_1)^{1/2}
\end{equation}
where $K$
is an elastic constant, $\eta$ a viscosity and $s_1$ is the shear
rate at the boundary\cite{deGennes93}.  
In terms of the parameters of our model this becomes
\begin{equation}
e_1\sim(\kappa \Gamma/ s_1)^{1/2}
\end{equation}
as the polymer viscosities can be shown to be
proportional to $1/\Gamma$ \cite{BE}. 

Similarly at the center of the flow the director must 
turn to extrapolate between the directions on the two sides.
This will happen over a distance
\begin{equation}
e_2\sim (\kappa \Gamma / s_2)^{1/2}.
\end{equation}
where $s_2$ is the shear rate at a distance $e_2/2$ from
the center.

For very low (or zero) flow rates the
boundary and central regions will join and the configuration (a)
with the director perpendicular to the flow at the center will be
preferred (this should be clear for the zero flow case where this
configuration has no elastic energy).  At high flow rates or for wider
channels the
configuration with the director aligned with the flow at the center will be
stable.  To see why this is so, consider the elastic
energy in the central region.  Outside the central region, the
director makes an angle $\pm \theta$ with the flow direction.  As can be
seen from equation~(\ref{shearangle}), this angle is always less than
$45^\circ$ and it will therefore cost less energy to go
through $0^\circ$ in the center than $90^\circ$.  Hence the flow
aligned configuration (b) is preferred.  
Note, however, that (a) can exist as a metastable state, given
suitable initial conditions.

\begin{figure}
\centerline{\epsfxsize=2.5in \epsffile{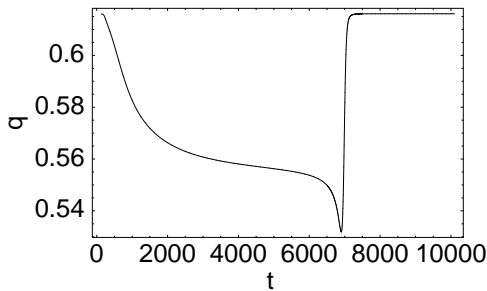}} \vskip
0.5true cm
\caption{The amplitude of the order parameter q at the center of the
  channel as a function of time $t$, measured in units of
  $1/(p_0\Gamma)$, for a flow strong enough to cause the system to switch
  between the two states (a) and (b) in Figure 1.  This occurs via 
  a two-stage process. The distorted region at the center is very 
slowly compacted, then quickly 'unzipped' by a defect running along
  the channel.}
\label{ampvtime}
\end{figure}

As the flow rate is increased the system
can spontaneously change the director
configurations at the center from perpendicular to parallel to the
flow.  This happens by nucleating a short
lived defect at the center which ``unzips'' the director.  
Figure~\ref{ampvtime} shows the amplitude of the order
parameter at the center of the flow as a function of time for a situation where
this happens. The sharp dip corresponds to the appearance of the defect.
To correctly
describe such a change of state clearly requires a formalism, such
as that used here, where the amplitude of the order parameter can
vary. 

\subsection{Shear thinning}

\begin{figure}
\centerline{\epsfxsize=2.5in \epsffile{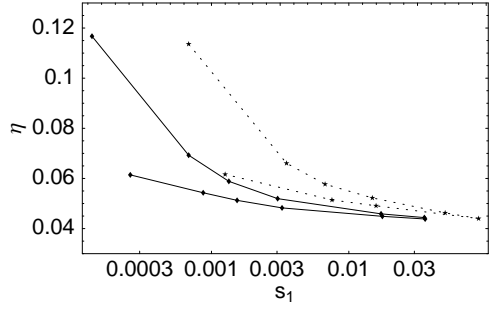}} \vskip
0.5true cm
\caption{Apparent viscosity $\eta$ as a function of the average strain.
  (The solid curves correspond to channels twice as wide as the dotted
  curves.)   
  Different branches correspond to the director perpendicular (upper
  branch) or parallel (lower branch) to the flow at the center of the
  channel.  The strain rate is scaled by $p_0\Gamma$ and the
  viscosity by $1/\Gamma$ to make them  dimensionless.}
\label{viscshear}
\end{figure}

The viscosity of a liquid crystal depends on its orientation with
respect to the flow \cite{deGennes93}.  As the orientation of the
director at the boundaries and center of the channel depends on the flow
rate the viscosity will be a function of the flow rate.
This dependence is shown in Figure~\ref{viscshear}. 
The apparent viscosity was calculated 
assuming equation(~\ref{upois}) holds (indeed the velocity
profile does remain close to parabolic).  This is equivalent to
the procedure used in experiments \cite{FF69} where equation (~\ref{flowrate})
is assumed to hold and the apparent viscosity is measured as a
function of flow rate.  

As expected the apparent viscosity also depends on the configuration
of the director at the center of the flow and hence the viscosity
curves in Figure 3 have two branches.  As the flow and hence the
strain rate is increased, the viscosity {\it decreases}: the fluid
exhibits shear thinning. 
At high shear rates the curves approach each other as the boundary and
central regions become smaller and the viscosity is
dominated by the flow-aligned regions.  

\begin{figure}
\centerline{\epsfxsize=2.5in \epsffile{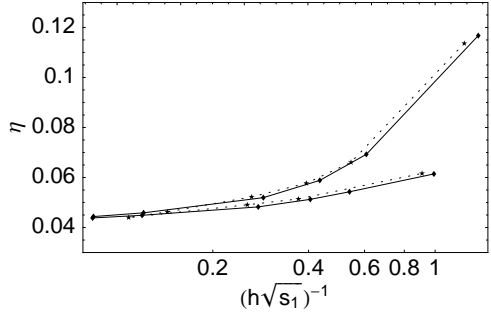}} \vskip
0.5true cm
\caption{Apparent viscosity as a function of the scaled variable 
  $(h\sqrt{s})^{-1}$. ($\kappa$ and the viscosity coefficients are held the
  same for the two cases, only the channel width is changed.)
  Different branches correspond to the director perpendicular (upper
  branch) or parallel (lower branch) to the flow at the center.  The
  units are the same as for Figure 3.}
\label{viscscale}
\end{figure}

Ericksen has shown \cite{E69}, based on a dimensional argument, that
the apparent viscosity at a shear rate $s$ can be written
\begin{equation}
\eta_{app}(s)=\eta_{app}(0) f\left[\frac{e_1}{h}\right]=\eta_{app}(0)
f\left[\left(\frac{\kappa \Gamma}{ s_1}\right)^{1/2}\frac{1}{h}\right],
\end{equation}
where the dimensionless function $f$ depends on the ratio between
various Leslie coefficients and the particular laminar flow under
study. Scaling the data
in this way, as shown in Figure~\ref{viscscale}, we find that the results
from the two different channel width do indeed collapse onto a single curve,
with two branches for the two different director configurations.   

\subsection{Log-rolling state}
\begin{figure}
\centerline{\epsfxsize=2.5in \epsffile{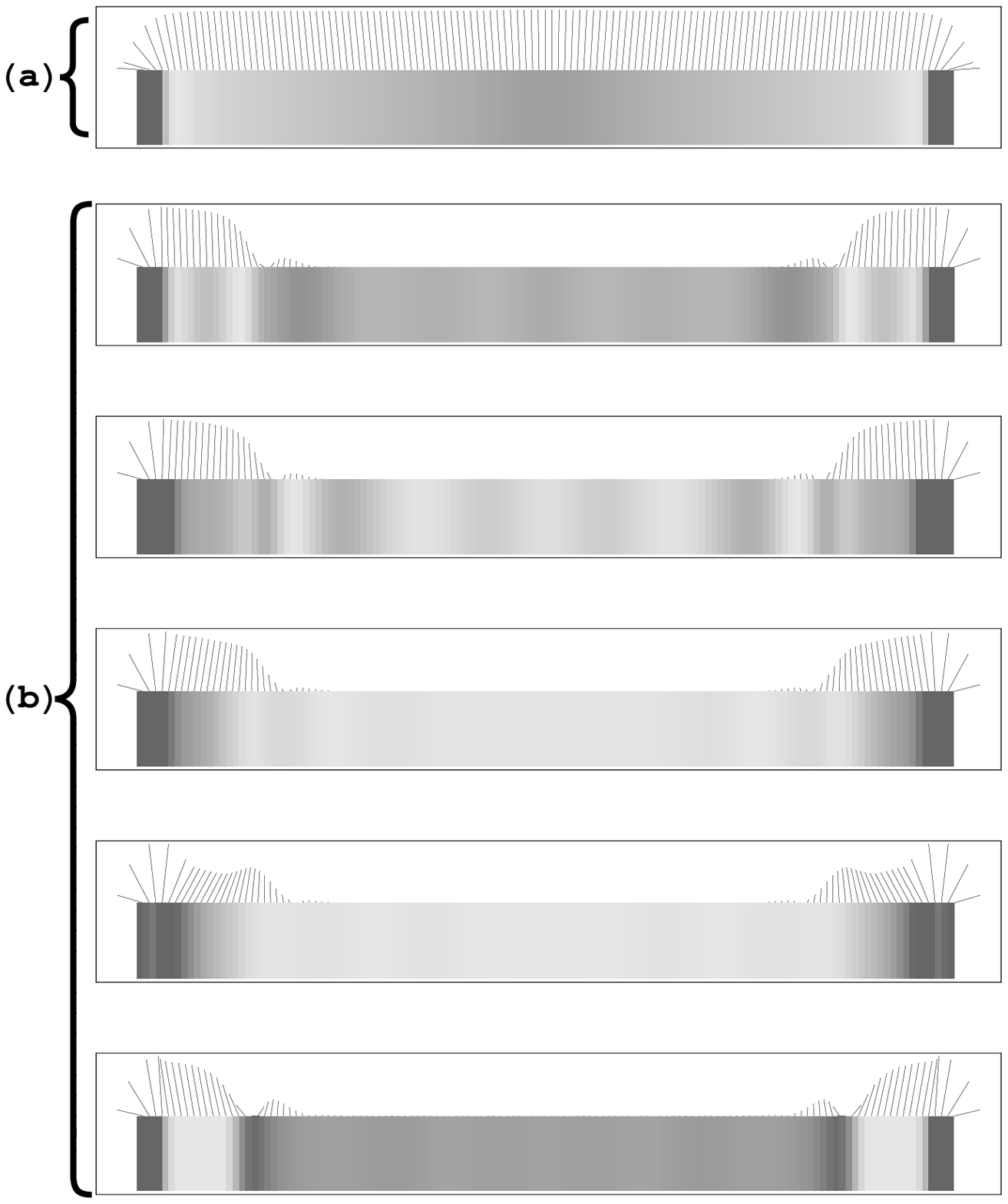}} \vskip
0.5true cm
\caption{Steady and oscillating states in Poiseuille flow.
  The lines represent the director orientation (eigenvector
  corresponding to largest eigenvalue of ${\bf Q}$) projected down onto the
  $xy$-plane, and shading represents the amplitude of the order
  parameter (largest eigenvalue).  Flow is from top to bottom, and the
  walls are at the left and right. At the walls, the director is aligned
  perpendicular to the boundary.  (a) Configuration (b) from Figure 1.  
  (b) Snapshots of an oscillating configuration  
  where the central region is in the ``log-rolling state'' (director
  perpendicular to the plane) and the boundary region consists of a
  transition from a configuration in the shear plane to a   
``tumbling'' (director rotating in the shear plane) and ``kayaking''
  region (director rotating in and out of the plane) interfacing to
  the central log-rolling state.}
\label{LogPois}
\end{figure}

\begin{figure}
\centerline{\epsfxsize=2.5in \epsffile{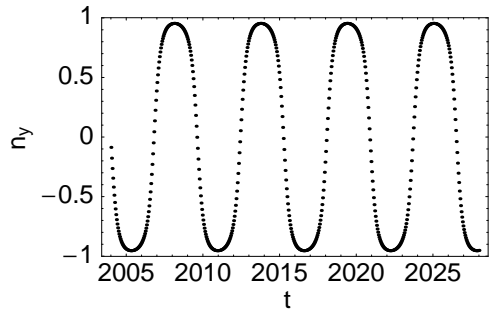}} \vskip
0.5true cm
\caption{Component of the director in the direction of fluid flow as a
  function of time at a point in the oscillating boundary layer (see
  Figure 5). }
\label{osc}
\end{figure}

So far we have assumed that the director stays in the shear plane.
This is not always the case.  If no solution to
equation (\ref{shearangle}) exists, the director tends to move out of
the plane to a log-rolling state.  In this state the director is
stationary and perpendicular to the shear plane.  However, the
eigenvector corresponding to the second largest eigenvalue of {\bf Q}
(the sub-director),
remains in the shear plane but oscillates (thus the ``rolling'' in
log-rolling).  As one approaches the boundary where the director is
constrained to be in the shear plane (and perpendicular to the wall)
the director must rotate back into the plane.   In this regime, the
director in the boundary layer may be stationary or may exhibit
oscillatory states \cite{TsujiRey}.  An example of one of these
dynamic flows is shown in Figure~\ref{LogPois}.  The oscillation of
the director in the boundary layer is plotted explicitly in
Figure~\ref{osc}.  There remain
interesting questions about the nature of the transitions between
different types of director behavior in shear flows.

\section{Discussion}

We have extended lattice Boltzmann algorithms of multiphase flow to
treat the case of a non-conserved, tensor order parameter. Hence it is
possible to simulate the Beris--Edwards equations of liquid crystal
hydrodynamics. These hold 
for the isotropic, uniaxial nematic, and biaxial nematic
phases.

Using the approach to investigate Poiseuille flow in liquid crystals a
rich phenomenology is apparent because of the coupling between the
director and the flow fields. In weak shear, for strong normal
anchoring at the boundaries, two states can be stabilized. For narrow
tubes the director field at the center of the flow prefers to align
perpendicular to the boundaries, for wider tubes it prefers to align
along the flow direction.
The effective viscosity decreases markedly with increases in the shear
rate. This is in agreement with the shear thinning seen in experiments
\cite{FF69}.

As the shear rate is increased the stable configuration changes to a
log-rolling state where the order parameter tumbles with the
flow. Oscillations in the director orientation in the 
boundary layer are also observed. In
experiments topological defects are nucleated at these shear rates,
perhaps because of impurities\cite{WF73,GM84}. 
In the simulations a similar phenomenon
may occur if fluctuations are added and work is in progress in this
direction.

There are many avenues for further research opened up by the rich
physics inherent in liquid crystal hydrodynamics and the generality of the
Beris--Edwards equations. The possibility of numerical
investigations will prove very helpful as the complexity of the
equations makes analytic progress difficult. Of particular interest
are the possibilities of exploring non-equilibrium phase transitions
such as shear banding: the coexistence of states with different strain
rates. In particular it will be possible to investigate the pathways
by which shear bands form. Work is also in progress to investigate the
effects of flow on the dynamics of topological defects.

\newpage

\end{document}